\documentclass[fleqn,twoside]{article}  
\usepackage[headings]{espcrc2}  
  
\readRCS  
$Id: espcrc2.tex,v 1.2 2004/02/24 11:22:11 spepping Exp $  
\usepackage{graphicx}  
\usepackage{amsmath}  
\usepackage[figuresright]{rotating}  
  
\title{Progress on NNLO subtraction\thanks{This research was supported   
in part by the Hungarian Scientific Research Fund (OTKA) grants T-038240   
and T-060432.}}  
  
\author{Vittorio Del Duca\address{INFN, sez. di Torino,  
via P. Giuria, 1 - 10125 Torino, Italy},  
G\'abor Somogyi\address{Institute of Nuclear Research of  
the Hungarian Academy of Sciences\\ and  
University of Debrecen, H-4001 Debrecen, PO Box 51, Hungary},   
Zolt\'an Tr\'ocs\'anyi\addressmark}

\runtitle{Progress on NNLO subtraction}  
\runauthor{V. Del Duca, G. Somogyi, Z. Tr\'ocs\'anyi}  
  
\begin{document}  
  
\begin{abstract}  
We review standard subtraction, as a method to compute  
cross sections at NNLO accuracy.  
\vspace{1pc}  
\end{abstract}  
  
\maketitle  
  
\section{Introduction}  
  
High-energy physics will enter a new era of discovery  
with the start of LHC  
operations in 2007. The LHC is a proton-proton collider that will  
function at the highest energy ever attained in the laboratory,  and will probe  
a new realm of high-energy physics.  
The use of a high-energy hadron collider as a research tool makes   
substantial demands  upon the  theoretical   
understanding  and  predictive  power  of  QCD,  
the  theory  of  the strong  interactions within the Standard Model.  
  
At high $Q^2$ any production rate can be expressed as a series expansion  
in $\alpha_S$. Because QCD is asymptotically free, the simplest approximation  
is to evaluate any series expansion to leading order in $\alpha_S$.  
However, for most processes a leading-order evaluation yields unreliable   
predictions. The next simplest approximation is the inclusion of  
radiative corrections at the NLO accuracy, which  
usually warrants a satisfying assessment of the production rates.  
In the previous decade, a lot of effort was devoted  
to devise process-independent   
methods and compute rates to NLO accuracy.  
In some cases, though, the NLO corrections may not be accurate enough.  
Specimen cases are: the extraction of $\alpha_S$ from fitting the  
predictions to the data, where   
in order to avoid that the main source of uncertainty be due to the   
NLO evaluation of some production rates, like the event shapes of jet  
production in $e^+e^-$ collisions, only observables evaluated to  
next-to-next-to-leading order (NNLO)  
accuracy are considered~\cite{Bethke:2004uy};   
open $b$-quark production at the Tevatron,  
where the NLO uncertainty bands are too large to test the   
theory~\cite{Cacciari:2003uh} {\it vs.} the data~\cite{Acosta:2004yw};  
Higgs production from gluon fusion in hadron collisions, where it is known  
that the NLO corrections are large~\cite{Graudenz:1992pv,Spira:1995rr},  
while the NNLO   
corrections~\cite{Harlander:2002wh,Anastasiou:2002yz,Ravindran:2003um},   
which have been evaluated in the large-$m_t$ limit,  
display a modest increase, of the order of less than 20\%, with respect to   
the NLO evaluation; Drell-Yan productions of $W$ and $Z$ vector bosons at LHC,  
which can be used as ``standard candles'' to measure the parton luminosity  
at the LHC~\cite{Dittmar:1997md,Khoze:2000db,Giele:2001ms,Frixione:2004us}.  
  
In the last few years the NNLO corrections have been computed to the  
total cross section~\cite{Harlander:2002wh,Hamberg:1990np} and to the  
rapidity distribution~\cite{Anastasiou:2003yy,Anastasiou:2003ds} of  
Drell-Yan production, to the total cross section for the production of  
a scalar~\cite{Harlander:2002wh,Anastasiou:2002yz,Ravindran:2003um} and  
a pseudoscalar~\cite{Harlander:2002vv,Anastasiou:2002wq} Higgs from  
gluon fusion as well as to a fully differential cross  
section~\cite{Anastasiou:2004xq,Anastasiou:2005qj}, and to jet  
production in $e^+e^-$ collisions~\cite{Anastasiou:2004qd,%  
Gehrmann-DeRidder:2004tv,Gehrmann-DeRidder:2005cm}.  
The methods which have been used are  
disparate: analytic integration, which is the first method to have been  
used~\cite{Hamberg:1990np}, cancels the divergences analytically, and  
is flexible enough to include a limited class of acceptance cuts by  
modelling cuts as propagators~\cite{Anastasiou:2002yz,Anastasiou:2003yy,%  
Anastasiou:2003ds,Anastasiou:2002wq};  
sector decomposition~\cite{Anastasiou:2004xq,Anastasiou:2004qd,%  
Roth:1996pd,Binoth:2000ps,Heinrich:2002rc,Gehrmann-DeRidder:2003bm,%  
Anastasiou:2003gr},   
which is flexible enough to include any acceptance cuts and for which  
the cancellation of the divergences is performed numerically;  
subtraction~\cite{Gehrmann-DeRidder:2005cm,Gehrmann-DeRidder:2003bm,%  
Kosower:2002su,Weinzierl:2003fx,Weinzierl:2003ra,Frixione:2004is,%  
Gehrmann-DeRidder:2005hi,Gehrmann-DeRidder:2005aw,Somogyi:2005xz},  
for which the cancellation of the divergences is  
organised in a process-independent way by exploiting the universal  
structure of the infrared divergences of a gauge theory, in particular  
the universal structure of the three-parton tree-level splitting  
functions~\cite{Berends:1988zn,Gehrmann-DeRidder:1997gf,Campbell:1997hg,%
Catani:1998nv,Catani:1999ss,DelDuca:1999ha}  
and the two-parton one-loop splitting   
functions~\cite{Bern:1994zx,Bern:1998sc,Kosower:1999xi,Kosower:1999rx,%
Bern:1999ry}.   
  
The standard approach of subtraction to NNLO relies on defining approximate  
cross sections which match the singular behaviour of the QCD cross  
sections in all the relevant unresolved limits.   
For processes without coloured partons in the initial state,  
we constructed in~\cite{Somogyi:2005xz} subtraction terms  
which regularise the kinematical singularities  
of the squared matrix element in all singly- and doubly-unresolved  
parts of the phase space. Thus, the  
regularised squared matrix element is integrable over all the phase  
space regions where at most two partons become unresolved.  
  
% Macros  
  
\def\beq{\begin{equation}}  
\def\eeq{\end{equation}}  
\def\beeq{\begin{eqnarray}}  
\def\eeeq{\end{eqnarray}}  
\def\aand{\!\!\!\!\!\!\!\!&&}  
  
\newcommand\bom[1]     {{\mbox{\boldmath $#1$}}}  
\newcommand\Refs[1]    {Refs.\,\cite{#1}}  
\newcommand\eqn[1]     {Eq.\,(\ref{#1})}  
\newcommand\eqns[2]    {Eqs.\,(\ref{#1}) and~(\ref{#2})}  
\newcommand\eqnss[2]   {Eqs.\,(\ref{#1})--(\ref{#2})}  
\newcommand\nn         {\nonumber}  
\newcommand\as         {\ensuremath{\alpha_{\mathrm{s}}}}  
\newcommand{\bT}       {\bom{T}}  
\newcommand{\eps}      {\varepsilon}          
\newcommand{\rd}       {{\mathrm{d}}}  
\newcommand\tsig[1]    {\sigma^{\mathrm{#1}}}  
\newcommand\dsig[1]    {\rd\sigma^{{\rm #1}}}  
\newcommand\dsiga[2]   {\rd\sigma^{{\rm #1,A}_{\scriptscriptstyle #2}}}  
\newcommand\M[2]       {\ensuremath{|{\cal{M}}_{#1}^{#2}|^2}}  
\newcommand\bra[3]     {\la {\cal M}_{#1}^{#2}#3|}  
\newcommand\ket[3]     {|{\cal M}_{#1}^{#2}#3\ra}  
\newcommand{\bA}[1]    {\bom{\mathrm A}_{#1}}  
\newcommand{\hP}       {\hat{P}}  
\newcommand{\cS}       {{\cal S}}  
\newcommand{\cC}[1]    {{\cal C}_{#1}}  
\newcommand{\cCS}[1]   {{\cal C}\kern-2pt{\cal S}_{#1}}  
\newcommand{\PS}[1]    {\rd\phi^{(#1)}}  
\newcommand{\ti}[1]    {\tilde{#1}}  
\newcommand{\wti}[1]   {\widetilde{#1}}  
\newcommand\tz[1]      {\tilde z_{#1}}  
\newcommand\tzz[2]     {\tilde z_{#1,#2}}  
\newcommand\kT[1]      {k_{\perp,#1}}  
\newcommand\kTt[1]     {\tilde{k}_{\perp,#1}}  
\newcommand\la         {\langle}  
\newcommand\ra         {\rangle}  
\newcommand{\IcC}[1]   {{\mathrm C}_{#1}}  
\newcommand{\IcS}[1]   {{\mathrm S}_{#1}}  
\newcommand{\qb}       {\bar{q}}  
\newcommand{\CF}       {C_{\mathrm{F}}}  
\newcommand{\CA}       {C_{\mathrm{A}}}  
\newcommand{\TR}       {T_{\mathrm{R}}}  
\newcommand\aeps       {\ensuremath{a_{\eps}}}  
\newcommand\Oe[1]      {\ensuremath{\mathrm O(\eps^{#1})}}  
\newcommand\bquad      {\!\!\!\!\!\!\!\!\!\!}  
\newcommand\Li{\mathop{\mathrm{Li}}\nolimits}  
  
\section{Subtraction scheme at NNLO}  
  
The NNLO correction to any $m$-jet cross section reads  
%is  
%a sum of three contributions, the doubly-real, the one-loop  
%singly-unresolved real-virtual and the two-loop doubly-virtual terms,  
\beeq  
\tsig{NNLO}   
\!\!\!  
&=&  
\!\!\!  
\int_{m+2}\!\dsig{RR}_{m+2} J_{m+2}  
+ \int_{m+1}\!\dsig{RV}_{m+1} J_{m+1}  
\nn\\  
&+& \!\!\!  
\int_m\!\dsig{VV}_m J_m\:.  
\label{eq:sigmaNNLO}  
\eeeq  
The three integrals in \eqn{eq:sigmaNNLO} are separately divergent,   
but their sum is finite for infrared-safe observables.  As explained in   
\cite{Somogyi:2005xz}, we rewrite \eqn{eq:sigmaNNLO} as  
\beeq  
\tsig{NNLO} \!\!\!&=&\!\!\!  
\int_{m+2}\!\dsig{NNLO}_{m+2}  
+ \int_{m+1}\!\dsig{NNLO}_{m+1}  
\nn\\ \qquad  
&+& \!\!\!   
\int_m\!\dsig{NNLO}_m\:,  
\label{eq:sigmaNNLOfin}  
\eeeq  
where each integral in \eqn{eq:sigmaNNLOfin} is finite by construction.  
Here we will focus on the subtractions that regularise doubly real  
emission, so we recall only that  
\beeq  
\dsig{NNLO}_{m+2}   
\!\!\!  
&=&  
\!\!\!  
\Bigg[\dsig{RR}_{m+2} J_{m+2} - \dsiga{RR}2_{m+2} J_m   
\label{eq:sigmaNNLOm+2}  
\\   
&-& \!\!\!  
\dsiga{RR}1_{m+2} J_{m+1} + \dsiga{RR}{12}_{m+2} J_m \Bigg]_{\eps =0}\:,  
\nn  
\eeeq  
where $\dsiga{RR}2_{m+2}$ and $\dsiga{RR}1_{m+2}$ regularise the doubly-   
and singly-unresolved limits of $\dsig{RR}_{m+2}$  respectively and  
$\dsiga{RR}{12}_{m+2}$ accounts for their overlap.

\iffalse  
where the integrands  
\beeq  
\dsig{NNLO}_{m+2}   
\!\!\!  
&=&  
\!\!\!  
\Bigg[\dsig{RR}_{m+2} J_{m+2} - \dsiga{RR}2_{m+2} J_m   
\label{eq:sigmaNNLOm+2}  
\\   
&-& \!\!\!  
\dsiga{RR}1_{m+2} J_{m+1} + \dsiga{RR}{12}_{m+2} J_m \Bigg]_{\eps =0}\:,  
\nn  
\eeeq  
\beeq  
\dsig{NNLO}_{m+1}   
\!\!\!  
&=&  
\!\!\!  
\Bigg[\dsig{RV}_{m+1} J_{m+1} - \dsiga{RV}{1}_{m+1} J_m  
\label{eq:sigmaNNLO3}  
\\ \qquad  
&& \!\!\!\!\!\!\!\!\!\!\!\!\!\!\!\!\!\!\!\!   
+ \int_1\!\Bigg(\dsiga{RR}1_{m+2} J_{m+1}  
 - \dsiga{RR}{12}_{m+2} J_m \Bigg) \Bigg]_{\eps =0}\:,  
\nn  
\eeeq  
and  
\beeq  
\dsig{NNLO}_m \!\!\!&=&\!\!\!  
\Bigg[\dsig{VV}_m J_m + \int_2\!\dsiga{RR}2_{m+2} J_m  
\label{eq:sigmaNNLO2}\\  
&+& \!\!\!   
\int_1\!\dsiga{RV}{1}_{m+1} J_m \Bigg]_{\eps = 0}  
\:,  
\nn  
\eeeq  
are sparately finite, thus integrable in four dimensions by  
construction.    
Above $\dsiga{RR}2_{m+2}$ and $\dsiga{RR}1_{m+2}$  
regularise the doubly- and singly-unresolved limits of $\dsig{RR}_{m+2}$   
respectively. Their overlap is added back in $\dsiga{RR}{12}_{m+2}$.  
Finally $\dsiga{RV}1_{m+1}$ is the counterterm regularising the  
singly-unresolved limits of $\dsig{RV}_{m+1}$. Here we focus on  
the subtractions that regularise doubly-real emission.  
\fi

\section{Subtractions for doubly-real emission}  
  
%\subsection{The general setup}  
  
The cross section $\dsig{RR}_{m+2}$, is the integral of the tree-level  
squared matrix element for ($m+2$)-parton production over the ($m+2$)-parton   
phase space  
\beq  
\dsig{RR}_{m+2} = \PS{m+2}{}\M{m+2}{(0)}\,.  
\label{eq:tsigRRm+2}  
\eeq  
The counterterms may be written symbolically as  
\beeq  
\dsiga{RR}2_{m+2} &=& \PS{m}{}\:[\rd p^{(2)}]{\bom{\cal A}}_2 \M{m+2}{(0)}\,,  
\label{eq:dsigRRA2}  
\\  
\dsiga{RR}1_{m+2} &=& \PS{m+1}{}\:[\rd p^{(1)}]{\bom{\cal A}}_1 \M{m+2}{(0)}\,,  
\label{eq:dsigRRA1}  
\eeeq  
and  
\beq  
\dsiga{RR}{12}_{m+2} =
\PS{m}{}\:[\rd p^{(1)}]\:[\rd p^{(1)}]{\bom{\cal A}}_{12}
\M{m+2}{(0)}\,.
\label{eq:dsigRRA12}  
\eeq
In this contribution we define explicitly the
${\bom{\cal A}}_1 \M{m+2}{(0)}$ term,
the terms ${\bom{\cal A}}_2 \M{m+2}{(0)}$
and ${\bom{\cal A}}_{12} \M{m+2}{(0)}$  
will be presented elsewhere.
  
The singly-singular subtraction ${\bom{\cal A}}_{1} \M{m+2}{(0)}$ is  
\beq  
{\bom{\cal A}}_{1}\M{m+2}{(0)} = \!\!\!  
\sum_{\stackrel{{\scriptstyle i,r}}{i\ne r}}  
 \left(\frac{1}{2} \cC{ir} - \cC{ir}\cS_{r}\right) + \sum_{r}\cS_{r}.  
\label{eq:A1}  
\eeq  
  
The singly-collinear term in \eqn{eq:A1} reads   
\beq  
\cC{ir} =   
8\pi\as\mu^{2\eps}\frac{1}{s_{ir}}  
\bra{m+1}{(0)}{}  
\hP_{f_i f_r}^{(0)}  
\ket{m+1}{(0)}{}\,,  
\label{eq:Cir}  
\eeq  
where $\hP_{f_i f_r}^{(0)} = \hP_{f_i f_r}^{(0)}(\tzz{i}{r},\tzz{r}{i},\kTt{ir};\eps)$ is   
the Altarelli-Parisi splitting function. The momentum fractions $\tzz{i}{r}$   
and $\tzz{r}{i}$ are defined as  
\beq  
\tzz{i}{r} = \frac{s_{iQ}}{s_{(ir)Q}}  
\quad\mbox{and}\quad  
\tzz{r}{i} = \frac{s_{rQ}}{s_{(ir)Q}}\,,  
\label{eq:ztirztri}  
\eeq  
and the transverse momentum $\kTt{ir}$ is given by  
\beq  
\kTt{ir}^{\mu} = \zeta_{i,r} p_r^{\mu} - \zeta_{r,i} p_i^{\mu}  
+(\tzz{r}{i}-\tzz{i}{r})\ti{p}_{ir}^{\mu}\,,  
\label{eq:kTtir}  
\eeq  
with  
\beq  
\zeta_{i,r} = \tzz{i}{r}-\frac{s_{ir}}{\alpha_{ir}s_{(ir)Q}}  
%\quad\mbox{and}\quad  
\, , \,  
\zeta_{r,i} = \tzz{r}{i}-\frac{s_{ir}}{\alpha_{ir}s_{(ir)Q}}.  
\label{eq:zeta}  
\eeq  
We used the abbreviations $s_{iQ}=2p_i\cdot Q$, $s_{rQ}=2p_r\cdot Q$   
and $s_{(ir)Q}=s_{iQ}+s_{rQ}$ above.  
  
The $m+1$ momenta entering the matrix elements on the right hand side of  
\eqn{eq:Cir} are  
\beeq  
\ti{p}_{ir}^{\mu} = \frac{p_i^{\mu} + p_r^{\mu} - \alpha_{ir} Q^{\mu}}{1-\alpha_{ir}}\,,  
\quad  
\ti{p}_n^{\mu} = \frac{p_n^{\mu}}{1-\alpha_{ir}}\,.  
%\qquad n\ne i,r\,,  
\label{eq:PS_single_coll}  
\eeeq  
In \eqn{eq:PS_single_coll} $n\ne i,r$ and  
\beq  
\alpha_{ir} =  
\frac{s_{(ir)Q}-\sqrt{(s_{(ir)Q})^2 - 4s_{ir}\: s}}{2s}  
\label{eq:alphair}  
\eeq  
with $Q^\mu$ the total four-momentum of the incoming electron and  
positron and $s = Q^2$.   
%Cleraly, the total four-momentum is conserved,  
%\beq  
%\ti{p}_{ir}^{\mu} + \sum_n \ti{p}_n^{\mu}  
%= p_i^\mu + p_r^\mu + \sum_n p_n^\mu\,.  
%\eeq  
  
The singly-soft term is  
\beq  
\cS_r =   
-8\pi\as\mu^{2\eps}  
\sum_{\stackrel{{\scriptstyle i,k}}{i\ne k}}  
\frac12 \cS_{ik}(r)\M{m+1,(i,k)}{(0)}\,,  
\label{eq:Sr}  
\eeq  
if $r$ is a gluon, and $\cS_r = 0$ if $r$ is a quark or antiquark.  
The $m+1$ momenta entering the matrix element on the right hand side of  
\eqn{eq:Sr} are  
%defined in by first rescaling all the hard momenta by a  
%factor $1/\lambda_r$ and then transforming all of the rescaled  
%momenta by a Lorentz transformation $\Lambda^{\mu}_{\nu}$,  
\beq  
\ti{p}_n^{\mu} = \Lambda^{\mu}_{\nu}[Q,(Q-p_r)/\lambda_r] (p_n^{\nu}/\lambda_r)\,,  
\quad n\ne r\,,  
\label{eq:PS_single_soft}  
\eeq  
where $\lambda_r=\sqrt{1-s_{rQ}/s}$ and  
\beq  
\Lambda^{\mu}_{\nu}[K,\wti{K}]=  
g^{\mu}_{\nu}  
- \frac{2(K+\wti{K})^{\mu}(K+\wti{K})_{\nu}}{(K+\wti{K})^{2}}   
+ \frac{2K^{\mu}\wti{K}_{\nu}}{K^2}\,.  
\label{eq:LambdaKKt}  
\eeq  
The matrix $\Lambda^{\mu}_{\nu}[K,\wti{K}]$ generates a (proper) Lorentz  
transformation, provided $K^2 = \wti{K}^2 \ne 0$.   
%Since $p_r^\mu$ is  
%massless ($p_r^2 = 0$), the total four-momentum is again conserved.  
   
In \eqn{eq:Sr} $\cS_{ik}(r) $ denotes the eikonal factor  
\beq  
\cS_{ik}(r) = \frac{2 s_{ik}}{s_{ir} s_{rk}}\,,  
\label{eq:Sikr}  
\eeq  
and the sum in \eqn{eq:Sr} runs over the external partons of the ($m+1$)-parton   
matrix element.  
  
The soft-collinear subtraction is defined by  
\beq  
\cC{ir}\cS_{r} =   
8\pi\as\mu^{2\eps} \frac{1}{s_{ir}}\frac{2\tzz{i}{r}}{\tzz{r}{i}}\,\bT_i^2\,  
\M{m+1}{(0)}\,,  
\label{eq:CirSr}  
\eeq  
if $r$ is a gluon, and $\cC{ir}\cS_{r} = 0$ if $r$ is a quark or antiquark.   
The momentum fractions are given by \eqn{eq:ztirztri}. The correct variables  
in the matrix element in the soft-collinear limit are those that appear in the  
soft limit \cite{Somogyi:2005xz}. Thus the $m+1$ momenta entering the matrix   
elements on the right hand side are given by \eqn{eq:PS_single_soft}.  
  
The momentum mappings introduced in \eqns{eq:PS_single_coll}{eq:PS_single_soft}  
in addition to conserving total four-momentum, both lead to exact phase-space   
factorisation in the form  
\beq  
\PS{m+2}=\PS{m+1} \: [\rd p^{(1)}]\,,  
\label{eq:PSfact}  
\eeq  
where the $m+1$ momenta in $\PS{m+1}$ are precisely  
those defined in \eqn{eq:PS_single_coll} or \eqn{eq:PS_single_soft}.   
  
The explicit expressions for $[\rd p^{(1)}]$ read  
%\beeq  
%~[\rd p^{(1)}] \aand=  
%\frac{(1-\alpha_{ir})^{m(d-2)-1}s_{\wti{ir}Q}}  
%{\sqrt{(s_{r\wti{ir}}+s_{\wti{ir}Q}-s_{rQ})^2+4s_{r\wti{ir}}(s-s_{\wti{ir}Q})}}  
%\,\Theta(1-\alpha_{ir})  
%\,\frac{\rd^d p_r}{(2\pi)^{d-1}}\delta_{+}(p_r^2)\,,  
%\label{eq:dp_coll}  
%\\   
%~[\rd p^{(1)}] \aand=  
%\lambda_{r}^{m(d-2)-2}\,\Theta(\lambda_{r})  
%\,\frac{\rd^d p_r}{(2\pi)^{d-1}}\delta_{+}(p_r^2)\,,  
%\label{eq:dp_soft}  
%\eeeq  
%  
\beq  
[\rd p^{(1)}] = {\cal J}\,\frac{\rd^d p_r}{(2\pi)^{d-1}}\delta_{+}(p_r^2)\,,  
\label{eq:dp1}  
\eeq  
and the Jacobian factors are  
\beeq  
{\cal J} \!\!\!\!\! &=& \!\!\!\!\!   
\frac{(1-\alpha_{ir})^{m(d-2)-1}\,\Theta(1-\alpha_{ir})\,s_{\wti{ir}Q}}  
%{\sqrt{(s_{r\wti{ir}}+s_{\wti{ir}Q}-s_{rQ})^2+4s_{r\wti{ir}}(s-s_{\wti{ir}Q})}}  
{\sqrt{(s_{r\wti{ir}}+s_{\wti{ir}Q}-s_{rQ})^2+4s_{r\wti{ir}}(s-s_{\wti{ir}Q})}}  
\label{eq:J_coll}  
\\  
{\cal J} \!\!\!\!\! &=& \!\!\!\!\! \lambda_{r}^{m(d-2)-2}\,\Theta(\lambda_{r})  
\label{eq:J_soft}  
\eeeq  
for the collinear and soft phase-space factorisations of  
\eqns{eq:PS_single_coll}{eq:PS_single_soft} respectively.  
  
In  
\eqn{eq:J_coll} $\alpha_{ir}$ is understood to be expressed in terms  
of the variable $\ti{p}_{ir}$,  
\beeq  
\alpha_{ir} \!\!\!\!\! &=& \!\!\!\!\!   
\Big[\sqrt{(s_{r\wti{ir}}+s_{\wti{ir}Q}-s_{rQ})^2+4s_{r\wti{ir}}(s-s_{\wti{ir}Q})}  
\nn\\ &-& \!\!\!\!\!  
(s_{r\wti{ir}}+s_{\wti{ir}Q}-s_{rQ})\Big]\Big[2(s-s_{\wti{ir}Q})\Big]^{-1}  
\label{eq:alpha_new}  
\eeeq  
  
The analytical integration of the counterterms over the factorised  
one-parton phase-space $[\rd p^{(1)}]$ is now possible.   
Starting with the collinear subtraction, notice that   
$\kTt{ir}$ as defined by \eqn{eq:kTtir} is orthogonal to $\ti{p}_{ir}$,  
therefore, the spin correlations generally present in \eqn{eq:Cir}  
vanish after azimuthal integration \cite{Catani:1996vz}.   
Thus when  
evaluating the integral of the subtraction term $\cC{ir}$   
over the factorised phase space $[\rd p^{(1)}]$,   
we may replace the Altarelli--Parisi splitting functions $\hP^{(0)}_{f_i f_r}$   
by their azimuthally averaged counterparts $P^{(0)}_{f_i f_r}$, so  
\beq  
\int[\rd p^{(1)}]\:\cC{ir} =  
%\frac{\as}{2\pi}\frac{1}{\Gamma(1-\eps)}\left(\frac{4\pi\mu^2}{Q^2}\right)^{\eps}  
\aeps  
\IcC{ir}(y_{\wti{ir}Q};m,\eps)\,\bT_{ir}^2\,\M{m+1}{(0)},  
\label{eq:IcC}
\eeq  
where $y_{\wti{ir}Q} = 2\ti{p}_{ir}\cdot Q/Q^2$,  
\beq  
\aeps = \frac{\as}{2\pi}\frac{1}{\Gamma(1-\eps)}\left(\frac{4\pi\mu^2}{Q^2}\right)^{\eps}\,  
\label{eq:aeps}  
\eeq  
and  
\beq  
%\frac{\as}{2\pi}\frac{1}{\Gamma(1-\eps)}\left(\frac{4\pi\mu^2}{Q^2}\right)^{\eps}  
\aeps  
\IcC{ir}(y_{\wti{ir}Q};m,\eps) =   
8\pi\as\mu^{2\eps}\!\int  
\frac{[\rd p^{(1)}]}{s_{ir}}\frac{P^{(0)}_{f_i f_r}}{\bT_{ir}^2}  
\,.  
\label{eq:ICir}  
\eeq  
  
The function $\IcC{ir}(y_{\wti{ir}Q};m,\eps)$ depends on the momentum of the  
parent parton and the flavours of the daughter partons. Explicitly  
\beeq  
\IcC{qg}(x;m,\eps) \!\!\!\!\! &=& \!\!\!\!\!  
x^{-2\eps}  
\Big[\,2\Big({\rm I}^{(-1)}_m(x;\eps) - {\rm I}^{(0)}_m(x;\eps)\Big)  
\nn\\ &+& \!\!\!\!\!  
(1-\eps){\rm I}^{(1)}_m(x;\eps)\Big],  
\label{eq:ICqg}\\  
\IcC{q\qb}(x;m,\eps) \!\!\!\!\! &=& \!\!\!\!\!  
\frac{\TR}{\CA}  
x^{-2\eps}  
\Big[\,{\rm I}^{(0)}_m(x;\eps) - \frac{2}{1-\eps}  
\nn\\ &\times& \!\!\!\!\!  
\Big({\rm I}^{(1)}_m(x;\eps) - {\rm I}^{(2)}_m(x;\eps)\Big)  
\Big],  
\label{eq:ICqqb}  
\eeeq  
and  
\beeq  
\IcC{gg}(x;m,\eps) \!\!\!\!\! &=& \!\!\!\!\!  
2x^{-2\eps}  
\Big[\,2\Big({\rm I}^{(-1)}_m(x;\eps) - {\rm I}^{(0)}_m(x;\eps)\Big)  
\nn\\ &+& \!\!\!\!\!  
{\rm I}^{(1)}_m(x;\eps) - {\rm I}^{(2)}_m(x;\eps)\Big].  
\label{eq:ICgg}  
\eeeq  
The analytical formulae for the $I^{(k)}_m(x;\eps)$ functions   
involve Beta functions, the ${}_2\!F_1$ hypergeometric function   
as well as the first Appell function $F_1$ and a certain   
generalisation of the last, $F_1^{(2)}$, see Table \ref{tb:Ikdefs}.   
%  
% Added the full analytical formulae  
%  
\begin{table*}[htb]  
\renewcommand{\arraystretch}{1.5} % enlarge line spacing  
\caption{Definitions of the $I^{(k)}_m(x;\eps)$ functions.  
$(a)_m = \Gamma(m+a)/\Gamma(a)$ denotes the Pochhammer symbol.}  
\label{tb:Ikdefs}  
\begin{tabular}{l|l}  
\hline  
\hline  
& $B(-2\eps,1-\eps)B(-\eps,1-\eps)$   
\\   
$I^{(-1)}_m(x;\eps)$  
& $\times\, F_1^{(2)}(-2\eps,-\eps,2m(1-\eps)-1-2\eps,-2(m-1)(1-\eps);1-3\eps,1-2\eps;1-x,1)$   
\\   
& $+B(1-2\eps,1-\eps)B(1-\eps,1-\eps)$   
\\   
& $F_1^{(2)}(1-2\eps,1-\eps,2m(1-\eps)-1-2\eps,-2(m-1)(1-\eps);2-3\eps,2-2\eps;1-x,1)$  
\\  
\hline  
$I^{(0)}_m(x;\eps)$ &  
$B(1-\eps,1-\eps)B(-\eps,2m(1-\eps)){}_2\!F_1(2m(1-\eps)-1-2\eps,-\eps;2m(1-\eps)-\eps;1-x)$  
\\  
\hline  
$I^{(k)}_m(x;\eps)$ &   
${\displaystyle \sum_{i=0}^{k}\binom{k}{i}}B(k-i+1-\eps,1-\eps)B(i-\eps,2m(1-\eps)+k-i)$  
\\  
& $F_1(i-\eps,2m(1-\eps)-1-2\eps,k;2m(1-\eps)+k-\eps;1-x,-1)\,,\quad k=1,2$  
\\  
\hline  
\hline  
\multicolumn{2}{l}{\vspace{1ex}\rule{0pt}{5ex}  
$F^{(2)}_1(a_1,a_2,b_1,b_2;c_1,c_2;x_1,x_2) =  
{\displaystyle \sum_{m_1=0}^{\infty}\sum_{m_2=0}^{\infty}}  
\frac{{\displaystyle (a_1)_{m_1+m_2}(a_2)_{m_1+m_2}(b_1)_{m_1}(b_2)_{m_2}}}{{\displaystyle (c_1)_{m_1+m_2}(c_2)_{m_1+m_2}}}  
\frac{{\displaystyle x_1^{m_1}}}{{\displaystyle m_1!}}\frac{{\displaystyle x_2^{m_2}}}{{\displaystyle m_2!}}$  
%\vspace{1ex}  
}\\  
\hline  
\hline  
\end{tabular}  
\end{table*}  
The expansion of $I^{(k)}_m(x;\eps)$  
in powers of $\eps$ is performed using the techniques of  
\cite{Moch:2001zr,Weinzierl:2002hv} to obtain  
\beeq  
\IcC{qg}(x;m,\eps) \!\!\!\!\! &=& \!\!\!\!\!  
\Big[\frac{1}{\eps^2}+\frac{3}{2\eps}-\frac{2}{\eps}\ln(x)+\Oe{0}\Big],  
\label{eq:IcCqgexp}  
\\  
\IcC{q\qb}(x;m,\eps) \!\!\!\!\! &=& \!\!\!\!\!  
\frac{\TR}{\CF}\Big[-\frac{2}{3\eps}+\Oe{0}\Big],  
\label{eq:IcCqqbexp}  
\\  
\IcC{gg}(x;m,\eps) \!\!\!\!\! &=& \!\!\!\!\!  
\Big[\frac{2}{\eps^2}+\frac{11}{3\eps}-\frac{4}{\eps}\ln(x)+\Oe{0}\Big].  
\label{eq:IcCggexp}  
\eeeq  
The finite parts, not shown here, depend on $m$ and can be easily found
for any given $m$ using the program of Ref.~\cite{Weinzierl:2002hv}.
  
Integrating the soft subtraction, the color correlations of   
\eqn{eq:Sr} are still present in the integrated expression  
\beeq  
\int [\rd p^{(1)}] \cS_{r} \aand=  
\sum_{\stackrel{{\scriptstyle i,k}}{i\ne k}}  
\aeps\IcS{ik}(y_{\ti{i}\ti{k}},y_{\ti{i}Q},y_{\ti{k}Q};m,\eps)  
\nn \\ &&\times\,  
\M{m+1;(i,k)}{(0)}\,,  
\label{eq:IcS}  
\eeeq  
where  
\beeq  
&&\bquad\!\!
\aeps\IcS{ik}(y_{\ti{i}\ti{k}},y_{\ti{i}Q},y_{\ti{k}Q};m,\eps) =  
\nn \\ && \qquad\qquad  
-8\pi\as\mu^{2\eps} \int [\rd p^{(1)}] \frac12 \cS_{ik}(r)\,.  
\eeeq  
For $\IcS{ik}(y_{\ti{i}\ti{k}},y_{\ti{i}Q},y_{\ti{k}Q};m,\eps)$ we find  
\beeq  
&&\bquad\!\!    
\IcS{ik}(y_{\ti{i}\ti{k}},y_{\ti{i}Q},y_{\ti{k}Q};m,\eps) =  
%\\ && \qquad  
%\frac{2}{\eps} \frac{\Gamma^2(1-\eps)}{\Gamma(1-2\eps)}  
%\,B(-2\eps,m(1-\eps)+1)  
-\frac{m(1-\eps)(1-2\eps)}{\eps^2}  
\nn\\ && \qquad  
\times\, B(1-\eps,1-\eps)B(1-2\eps,m(1-\eps))  
\nn \\ && \qquad\times\,  
\kappa \,{}_2\!F_1(1,1,1-\eps,1-\kappa)\,,  
\label{eq:ISik}  
\eeeq  
where $\kappa = y_{\ti{i}\ti{k}}/(y_{\ti{i}Q}y_{\ti{k}Q})$ and the
hypergeometric function can be expanded using
\beeq  
&&\bquad\!\!
\kappa\;{}_2\!F_1(1,1,1-\eps,1-\kappa) =  
\nn\\&& \qquad  
\kappa^{-\eps}\,\Big[1 + \eps^2 \Li_2(1-\kappa) + \Oe{3}\Big]\,.  
\qquad~  
\eeeq  
%Fianlly integrating the soft-collinear subtraction, \eqn{eq:CirSr} we obtain  
  
The integral of the soft-collinear counterterm, \eqn{eq:CirSr}, reads  
\beq  
\int [\rd p^{(1)}] \cC{ir}\cS_{r} =   
\aeps\IcC{ir}\IcS{r}(m,\eps) \, \bT_i^2 \, \M{m+1}{(0)}  
\label{eq:IcCS}  
\eeq  
with  
\beq  
\aeps\IcC{ir}\IcS{r}(m,\eps) = 8\pi\as\mu^{2\eps}  
\int [\rd p^{(1)}] \frac{1}{s_{ir}}\frac{2\tzz{i}{r}}{\tzz{i}{r}}\,.  
\eeq  
Performing the integrations, we obtain
\beeq  
\IcC{ir}\IcS{r}(m,\eps) \!\!\! &=& \!\!\!  
%2\Big[1+\frac{m(1-\eps)(1-2\eps)}{2\eps^2}\Big]  
\Big[\frac{m(1-\eps)(1-2\eps)}{\eps^2} + 2\Big]  
\nn\\ && \bquad\bquad\bquad  
\times\, B(1-\eps,1-\eps)B(1-2\eps,m(1-\eps))  
\,.  
\label{eq:ICS}  
\eeeq  
  
In order to show that the integrated subtraction term
$\int_1 {\bom{\cal A}}_{1} \M{m+2}{(0)}$   has the correct pole
structure \cite{Catani:1996vz}, we exploit colour conservation in
\eqns{eq:IcC}{eq:IcCS}. Indeed, using $\bT_i^2 = -\sum_{k\ne i}\bT_i\bT_k$,
we can combine the soft and soft-collinear subtractions in
\eqns{eq:IcS}{eq:IcCS},
\beeq  
&&\bquad\!\!  
\int_1\Big(\sum_{r}\cS_{r}  
-\sum_{\stackrel{{\scriptstyle i,r}}{i\ne r}}\cC{ir}\cS_{r}\Big) =  
\\ && \qquad  
\aeps \sum_{r}\sum_{\stackrel{{\scriptstyle i,k}}{i\ne k}}  
\Big(\IcS{ik} + \IcC{ir}\IcS{r}\Big) \M{m+1,(i,k)}{(0)}\,,  
\nn  
\eeeq  
Expanding $(\IcS{ik} + \IcC{ir}\IcS{r})$ in powers of $\eps$, we find  
\beq  
\IcS{ik} + \IcC{ir}\IcS{r} = \frac{1}{\eps}  
\ln \frac{y_{\ti{i}\ti{k}}}{y_{\ti{i}Q} y_{\ti{k}Q}} + \Oe{0}\,.  
\label{eq:IcSCSexp}  
\eeq  
We can combine these contributions with the collinear functions in
\eqnss{eq:IcCqgexp}{eq:IcCggexp} after using colour conservation in
\eqn{eq:IcC} and find that the $\ln y_{\ti{i}Q}/\eps$ terms in
\eqnss{eq:IcCqgexp}{eq:IcCggexp} and \eqn{eq:IcSCSexp} exactly cancel
and the known structure of the one-loop squared matrix elements is
reproduced.

\section{Conclusion}  
  
We outlined a subtraction scheme for computing cross sections at  
NNLO accuracy using the known singly- and doubly-singular limits  
of squared matrix elements. A way of disentangling these overlapping  
limits was presented in \cite{Somogyi:2005xz}. Here we discussed how   
to define the singly-singular subtraction terms.  
We presented exact phase space factorisations that allow us the  
integration of the singular factors and the results of these integrations.  
We have also coded  
\eqn{eq:sigmaNNLOm+2} for the case when $\dsig{RR}_{m+2}$ is the fully  
differential cross section for the process $e^+e^- \to q \bar{q} g g g$  
($m=3$) and $J_5$ defines the $C$-parameter. We found that the  
integral of $\dsig{NNLO}_{m+2}$ is finite and integrable in four  
dimensions using standard Monte Carlo methods.

\iffalse  
The subtraction scheme outlined here uses the known singly- and  
doubly-singular limits of the squared matrix elements. These limits  
overlap and a way of disentanglement was presented in \cite{Somogyi:2005xz}.   
In this contribution we discussed how to make the next step, namely  
we outlined the exact phase space factorisations we propose for the  
collinear and soft subtraction terms.  Putting the subtraction terms on  
the factorised phase space allows us the integration of the singular  
factors such that the remaining expressions can be combined with the  
virtual correction. This integration and combination is left for future  
work.  
\fi

\end{document}